      \documentstyle[12pt,graphicx]{article}                
\begin{document}
\baselineskip 18pt
\def\today{\ifcase\month\or
 January\or February\or March\or April\or May\or June\or
 July\or August\or September\or October\or November\or December\fi
 \space\number\day, \number\year}
\def\thebibliography#1{\section*{References\markboth
 {References}{References}}\list
 {[\arabic{enumi}]}{\settowidth\labelwidth{[#1]}
 \leftmargin\labelwidth
 \advance\leftmargin\labelsep
 \usecounter{enumi}}
 \def\newblock{\hskip .11em plus .33em minus .07em}
 \sloppy
 \sfcode`\.=1000\relax}
\let\endthebibliography=\endlist
\def\lsim{\ ^<\llap{$_\sim$}\ }
\def\gsim{\ ^>\llap{$_\sim$}\ }
\def\r2{\sqrt 2}
\def\beq{\begin{equation}}
\def\eeq{\end{equation}}
\def\beqn{\begin{eqnarray}}
\def\eeqn{\end{eqnarray}}
\def\rmuu{\gamma^{\mu}}
\def\rmud{\gamma_{\mu}}
\def\PL{{1-\gamma_5\over 2}}
\def\PR{{1+\gamma_5\over 2}}
\def\sinW2{\sin^2\theta_W}
\def\AEM{\alpha_{EM}}
\def\mul{M_{\tilde{u} L}^2}
\def\mur{M_{\tilde{u} R}^2}
\def\mdl{M_{\tilde{d} L}^2}
\def\mdr{M_{\tilde{d} R}^2}
\def\mz2{M_{z}^2}
\def\c2b{\cos 2\beta}
\def\au{A_u}         
\def\ad{A_d}
\def\cob{\cot \beta}
\def\v#1{v_#1}
\def\tb{\tan\beta}
\def\epem{$e^+e^-$}
\def\KK{$K^0$-$\bar{K^0}$}
\def\wi{\omega_i}
\def\xj{\chi_j}
\def\Wmu{W_\mu}
\def\Wnu{W_\nu}
\def\m#1{{\tilde m}_#1}
\def\mH{m_H}
\def\mw#1{{\tilde m}_{\omega #1}}
\def\mx#1{{\tilde m}_{\chi^{0}_#1}}
\def\mc#1{{\tilde m}_{\chi^{+}_#1}}
\def\mwi{{\tilde m}_{\omega i}}
\def\mxi{{\tilde m}_{\chi^{0}_i}}
\def\mci{{\tilde m}_{\chi^{+}_i}}
\def\mz{M_z}
\def\sw{\sin\theta_W}
\def\cw{\cos\theta_W}
\def\cb{\cos\beta}
\def\sb{\sin\beta}
\def\rwi{r_{\omega i}}
\def\rxj{r_{\chi j}}
\def\rfp{r_f'}
\def\Kik{K_{ik}}
\def\Fq2{F_{2}(q^2)}
\def\f{\({\cal F}\)}
\def\d1{{\f(\tilde c;\tilde s;\tilde W)+ \f(\tilde c;\tilde \mu;\tilde W)}}
%%%%%%%%%%%%%%%%%%%%%%%%%%%%%%%%%%
\def\tw{\tan\theta_W}
\def\sec2w{sec^2\theta_W}
%%%%%%%%%%%%%%%%%%%%%%%%%%%%%%%%%%

\begin{titlepage}
%{\flushleft 
%{NUB-TH-}}
%%\hspace*{10.0cm} OCHA-PP-78
%%\hspace*{10.0cm} \today
%\vspace{1cm}
\begin{center}
{\large {\bf Effects of Large CP Phases on the Proton Lifetime in
Supersymmetric Unification }}\\
\vskip 0.5 true cm
\vspace{2cm}
\renewcommand{\thefootnote}
{\fnsymbol{footnote}}
 Tarek Ibrahim$^{a,b}$ and Pran Nath$^{b}$  
\vskip 0.5 true cm
\end{center}

\noindent
{a. Department of  Physics, Faculty of Science,
University of Alexandria,}\\
{ Alexandria, Egypt}\\ 
{b. Department of Physics, Northeastern University,
Boston, MA 02115-5000, USA } \\
\vskip 1.0 true cm
\centerline{\bf Abstract}
\medskip
The effects of large CP violating phases arising from the 
soft SUSY breaking parameters  on the proton lifetime are  investigated
in supersymmetric grand unified models.
It is found that the CP violating phases can reduce as well as 
enhance the proton lifetime depending on the part of the parameter space
 one is in. Modifications of the proton lifetime by as much as a factor
 of 2 due to the effects of the CP violating phases are seen.
 The largest effects arise for the lightest sparticle spectrum 
 in the dressing loop integrals and the effects decrease with 
 the increasing 
 scale of the sparticle masses. An analysis of the uncertainties in
 the determination of the proton life time due to uncertainties in
 the quark masses and in the other input data is also given.  
 These results  are  of  import in the precision predictions
 of the proton lifetime in supersymmetric unification 
 both in GUT and in string models when the soft SUSY breaking 
 parameters are complex. 
\end{titlepage}

\section{Introduction}
It is well known that there are new sources of CP violation 
in supersymmetric theories which arise from the soft SUSY 
breaking sector of the theory. The normal size of such phases
is O(1) and an order of magnitude estimate shows 
that such large phases
would lead to a conflict with the current experimental 
limits on
the electron\cite{commins} and on the neutron electric dipole 
moment\cite{harris}. The 
conventional ways suggested to avoid this conflict is either to
assume that the phases are small\cite{ellis,wein} or that the
SUSY spectrum is heavy\cite{na}. However,
recently it was demonstrated\cite{in1} that this need not 
to be 
the case and indeed
there could be consistency with experiment even with large CP 
violating phases and a light spectrum due to an internal 
cancellation mechanism among
the various contributions to the EDMs. The above possibility has
led to a considerable further activity
 \cite{in2,in4,pilaftsis} and the effects of large CP phases  under the
 cancellation mechanism have been investigated in
dark matter with the EDM constraints\cite{cin}, 
in $g_{\mu}-2$\cite{g2} and 
 in other low energy physics phenomena\cite{more}. 

 In this paper we investigate the effects of large CP violating phases
 on nucleon stability in supersymmetric 
 grand unification with  baryon and lepton number violating 
 dimension five operators\cite{wein1,acn,acn1}. 
 The main result of this analysis is that the dressing loop 
 integrals that enter in the supersymmetric proton decay analysis
 are modified due to the effect of the large CP violating phases.
 The CP effects on the proton life time are most easily exhibited
 by considering $R_{\tau}$ defined in Eq.(22) which is the ratio
 of the p lifetime with phases and without phases. $R_{\tau}$
 is largely independent of the GUT structure which cancels out
 in the ratio. Since the dressing loop integrals enter in the proton
 decay lifetime in both GUT and string models which contain  
 the baryon and the lepton number violating dimension five operators,
 the phenomena of CP violating effects on the proton lifetime should 
 hold  for a wide range of models  both of GUT and of 
 string variety\cite{string}.
 However, for concreteness we will consider first 
 the simplest SU(5) supersymmetric grand unified model, and then 
 consider a non-minimal extension. As discussed above 
 similar analyses should hold for a wider class of models and so
 what we do below should serve as an illustration of the general
 idea of the effect of large  CP phases on the proton lifetime.

 The outline of the paper is as follows: In Sec.2 we give 
 a theoretical analysis of the effects of CP violating 
 phases on proton decay in the minimal
  supersymmetric  SU(5) model for specificity.
  In Sec.3 we discuss the numerical effects of the CP violating
  phases on $R_{\tau}$ under the EDM constraints. A
  non-minimal extension is also discussed and an analysis of
  the uncertainties in the predictions of the proton life time
  due to uncertainties in the quark masses, in $\beta_p$ and in 
  the KM matrix elements is  given. Conclusions 
  are given in Sec.4.

\section{Theoretical analysis of CP violating  phases on 
proton decay in supersymmetric GUTs}
In the minimal 
supergravity unified model (mSUGRA)\cite{chams} 
the soft SUSY breaking
can be parameterized by $m_0$, $m_{\frac{1}{2}}$, $A_0$,
and $\tan\beta$, where $m_0$ is the universal scalar mass, 
$m_{\frac{1}{2}}$ is the universal gaugino mass, $A_0$ is the 
universal trilinear coupling all taken at the GUT scale, and
$\tan\beta=<H_2>/<H_1>$ is the ratio of the Higgs VEVs where
$H_2$ gives mass to the up quark and $H_1$ gives mass to the
down quark and the lepton. In addition, the effective theory 
below the  GUT scale contains the Higgs mixing parameter 
$\mu$ which enters in the superpotential in the term $\mu H_1H_2$.
In the presence of CP violation one finds that the minimal model
contains two independent CP violating phases  which can be taken
to be $\theta_{\mu}$, which is the phase of $\mu$ and $\alpha_{A_0}$
which is the phase of $A_0$.

For more general situations when one allows for non-universalities,
the soft SUSY breaking sector of the theory brings in more CP 
violating phases. Thus unlike the case of mSUGRA here  the 
$U(1)\times SU(2)\times SU(3)$ gaugino masses $\tilde m_i$
(i=1,2,3) can have arbitrary phases, i.e., 
\begin{equation}
\tilde m_i=|\tilde m_i|e^{i\xi_i}~~(i=1,2,3)
\end{equation} 
While in the universal case a field redefinition can eliminate
the common phase of the gaugino masses, here one finds that 
the difference of the gaugino phases does persist in the
low energy theory and in
fact is found to be a useful tool in 
arranging for the cancellation mechansim to work for the satisfaction
of the EDMs. In the following analysis  we carry out an
analysis of the proton decay with the most general allowed 
set of CP violating phases. The definition of the mass matrices
for charginos, neutralinos and for squarks  and sleptons have
been explicitly exhibited in Ref.\cite{in3} and we refer the reader 
to this paper for details. 
 The focus of the present work is to analyze the effects of CP 
 violating phases on p decay  and to estimate  its size.
 For the sake of concreteness we begin with a discussion of 
  the simplest grand
 unification model, i.e., the minimal SU(5) model. However, the 
 technique discussed here to include the CP effects on p decay 
 can be used to anlayse the CP violating effects for any 
 supersymmetric unified model with baryon and lepton number violating
 dimension five operators. 
 This class includes string models. 

  As mentioned above we consider for concreteness and simplicity   
   the minimal SU(5) model whose matter interactions are 
     given by\cite{wein1,acn,acn1}
  \beq
W_Y=-\frac{1}{8}f_{1ij}\epsilon_{uvwxy}H_1^uM_i^{vw}M_j^{xy}+
f_{2ij}\bar H_{2u}\bar M_{iv} M_j^{uv}
\eeq
where $M_u,M^{uv}$ are the $\bar 5, 10$  plet representations of
SU(5), and $H_1,H_2$ are the $\bar 5, 5$ of SU(5).
After the breakdown of the GUT symmetry and 
integration over the Higgs triplet fields the effective dimension
five interactions  below the GUT scale which governs p decay  
is given by\cite{wein1,acn,acn1}
\beq
{\cal L}_{5L}= \frac{1}{M} \epsilon_{abc}(Pf_1^uV)_{ij}(f_2^d)_{kl}
( \tilde u_{Lbi}\tilde d_{Lcj}(\bar e^c_{Lk}(Vu_L)_{al}-
\bar \nu^c_kd_{Lal})+...)+H.c.
\eeq
\beq
 {\cal L}_{5R}= -\frac{1}{M} \epsilon_{abc}(V^{\dagger} f^u)_{ij}(PVf^d)_{kl}
(\bar e^c_{Ri}u_{Raj}\tilde u_{Rck}\tilde d_{Rbl}+...)+H.c.
\eeq
where ${\cal L}_{5L}$ and ${\cal L}_{5R}$
 are the LLLL and RRRR  
lepton and baryon number violating dimension 5 operators,
 V is the CKM matrix and $f_i$ are  related to quark masses,
and $P_i$ appearing in Eqs.(3) and (4) are the generational phases
 given by $P_i=(e^{i\gamma_i})$, $\sum_i \gamma_i=0$  (i=1,2,3).

The baryon and the lepton number violating dimension five operators must
be dressed by the chargino, the gluino and the neutralino exchanges 
to generate 
 effective baryon and lepton number violating dimension six operators 
 at low energy (Some examples of dressing loop diagrams are given 
in Fig.1). It is in this process of dressing of the dimension
 five operators that the CP violating phases of the soft SUSY breaking
 sector enter in the proton decay amplitude. The CP phases enter the
dressings in two ways,  via the mass matrices of the charginos,
the neutralinos and the sfermions, and via the interaction vertices.  
Taking account of this additional complexity, the analysis
for computing the proton decay amplitudes follows the usual procedure. 
Thus to   
dress the dimension five operators the squark and slepton fields 
must be eliminated in terms of their sources. As an example, the
up squarks in the presence of CP violating phases can be eliminated
using the relations
\beqn
\tilde u_{iL}=2\int [\Delta_{ui}^L L_{ui}+\Delta_i^{LR} R_{ui}]\nonumber\\
\tilde u_{iR}=2\int [\Delta_{ui}^R R_{ui}+\Delta_i^{RL} L_{ui}]
\eeqn
where 
\beq
L_{ui}=\frac{\delta \it L_I}{\delta \tilde u_{iL}^{\dagger}},
R_{ui}=\frac{\delta \it L_I}{\delta \tilde u_{iR}^{\dagger}}
\eeq
Here $L_I$ is the sum of fermion-sfermion-gluino, 
fermion-sfermion-chargino and fermion-sfermion-neutralino 
interactions 
and 
\beqn
\Delta_{ui}^L= [|D_{ui11}|^2\Delta_{ui1}+|D_{ui12}|^2\Delta u_{i2}]\nonumber\\
\Delta_{ui}^R= [|D_{ui21}|^2\Delta_{ui1}+|D_{ui22}|^2\Delta u_{i2}]
\eeqn 
and 
\beqn
\Delta_{ui}^{LR}= -D_{ui11}D_{ui12}[\Delta_{ui1}-\Delta u_{i2}]\nonumber\\
\Delta_{ui}^{RL}= -D_{ui11}D_{ui12}^*[\Delta_{ui1}-\Delta u_{i2}].
\eeqn
 Here $\tilde u_{i1}$ and $\tilde u_{i2}$ are the squark mass
eigenstates for the squark flavors $u_{i1}$ and $u_{i2}$
and  $\Delta_{u_i1}$ and $\Delta_{u_i2}$ are the corresponding 
propagators, and
 $D_{ui}$ is the diagonalizing matrix for the $\tilde u_{i}$
squarks, i.e., 
\beq
D_{u_i}^{\dagger} M_{\tilde u_{i}}^2 D_{u_i}=diag(M_{\tilde u_{i1}}^2,
M_{\tilde u_{i2}}^2)
\eeq
We note the special arrangement of the complex quantities and their
complex conjugates in Eqs.7 and 8. Specifically we note that  
 while in the absence of CP phases 
$\Delta_{ui}^{LR}$ = $\Delta_{ui}^{RL}$ this is not the case in the
 presence of CP phases and 
 in general one has 
$\Delta_{ui}^{LR} \neq \Delta_{ui}^{RL}$ as is seen from
Eqs.(8). 
 $L_{ui}$ and $R_{ui}$ defined  by Eq.(6) receive 
contributions from the chargino, the neutralino and the gluino 
exchanges. 

  Following the standard procedure\cite{wein1,acn,acn1} one obtains
   the effective dimension six operators for the baryon and the lepton
  number violating interaction arising from dressing of the 
  dimension five operators. 
 From this effective interaction one obtains
 the proton lifetime decay widths for various modes using
 the effective Lagrangian methods. We 
 limit ourselves here to the dominant decay mode 
 $p\rightarrow \bar\nu_iK^+$. 
  Including the CP violating effects the decay width for this 
  process is given by
\beqn
\Gamma(p\rightarrow\bar\nu_iK^+)=\frac{\beta_p^2m_N}{M_{H_3}^232\pi f_{\pi}^2}
(1-\frac{m_K^2}{m_N^2})^2|{\cal A}_{\nu_iK}|^2 A_L^2(A_S^L)^2\nonumber\\
|(1+\frac{m_N(D+3F)}{3m_B})(1+{\cal Y}_i^{tk}+(e^{-i\xi_3}{\cal Y}_{\tilde g}+
{\cal Y}_{\tilde Z})\delta_{i2}+
\frac{A_S^R}{A_S^L}{\cal Y}_1^R\delta_{i3})\nonumber\\
+\frac{2}{3}\frac{m_N}{m_B}D(1+{\cal Y}_3^{tk}-
(e^{-i\xi_3}{\cal Y}_{\tilde g}-{\cal Y}_{\tilde Z})\delta_{i2}
+\frac{A_S^R}{A_S^L}{\cal Y}_2^R \delta_{i3})|^2
\eeqn
where
\beq 
{\cal A}_{\nu_iK}=(\sin 2\beta M_W^2)^{-1}\alpha_2^2P_2m_cm_i^dV_{i1}^{\dagger}
V_{21}V_{22}[{\cal F}(\tilde c;\tilde d_i;\tilde W)+
 {\cal F}(\tilde c;\tilde e_i;\tilde W)]
\eeq
In the above $A_L(A_S)$ are the long (short) suppression factors, 
D,F, $f_{\pi}$ are the effective Lagrangian parameters, and 
$\beta_p$ is defined by 
$\beta_p U_L^{\gamma}=\epsilon_{abc}\epsilon_{\alpha \beta} <0|d_{aL}^{\alpha}
u_{bL}^{\beta}u_{cL}^{\gamma}|p>$ where $U_L^{\gamma}$ is the 
proton wavefunction. Theoretical determinations of $\beta_p$ lie in the
range $0.003-0.03~GeV^3$. 
Perhaps the more reliable estimate is from lattice gauge calculations 
which gives\cite{gavela} 
$\beta_p=(5.6\pm 0.5)\times  10^{-3} GeV^3$. 

 Aside from the explicit CP  phases via the exponential factor
 $e^{-i\xi_3}$ in Eq.(10), CP effects
 enter dominantly in \f 's which are the dressing loop integrals.
 For the chargino exchange 
 in the presence of CP violating phases one has 
\beqn
{\cal F}(\tilde u_i;\tilde d_j; \tilde W)= -32\pi^2i \int 
\sum_{A=1,2}[\Delta^L_{uai}S_{A1}^*-\Delta_{uai}^{LR}\epsilon_i^u
S_{A2}^*]\nonumber\\
\tilde G_A[\Delta^L_{dj}U_{A1}^*-\Delta_{dj}^{LR}\epsilon_j^d
U_{A2}^*]
\eeqn
  Here $\tilde G_A$ (A=1,2) are the propagators for the chargino mass 
  eigenstates
  and the  matrices U and S enter in the biunitary 
  transformations to diagonlize the chargino mass matrix $M_C$ such that
  \beq
 U^*M_CS^{-1}=diag(\tilde m_{\chi_1^+},\tilde m_{\chi_2^+})
  \eeq
 In Eq.(10) the quantities ${\cal Y}_i^{tk}$ are the corrections due 
 to the chargino exchanges
 involving  third generation squarks, ${\cal Y}_{\tilde g}$  is the 
 contribution from the gluino exchange, ${\cal Y}_{\tilde Z}$ is the
 contribution from the neutralino exchange, and ${\cal Y}_i^R$ are the 
 contributions from the dressing of the RRRR dimension 5 operators.

 The gluino exchange contribution  ${\cal Y}_{\tilde g}$ is given by
\beq
 {\cal Y}_{\tilde g}=\frac{4}{3}\frac{P_1}{P_2}\frac{\alpha_3}{\alpha_2}
 \frac{m_u V_{11}}{m_cV_{21}V_{21}^{\dagger}V_{22}}
\frac{{\cal H}(\tilde u;\tilde d;\tilde g)-{\cal H}(\tilde d;\tilde d;\tilde g)}
 {{\cal F}(\tilde c;\tilde s;\tilde W)+{\cal F}(\tilde c;\tilde \mu;\tilde W)}
 \eeq
The function ${\cal H}$ is defined by ${\cal H}(\tilde u;\tilde d;\tilde g)
=f(\tilde m_u; \tilde m_d; \tilde m_g)$ where f is defined by
Eq.(19) below. The contributions ${\cal Y}_i^R$ from the dresssing of the RRRR 
dimension five  operators are given by
\beq
{\cal Y}_1^R= \frac{P_1}{P_2} \frac{m_tm_dV_{11}V_{32}V_{33}^{\dagger}}
{m_cm_bV_{21}V{22}V_{31}^{\dagger}}
\frac{{\cal Q}(\tilde \tau;\tilde t;\tilde W)}
{{\cal F}(\tilde c;\tilde b;\tilde W)+{\cal F}(\tilde c;\tilde \tau;\tilde W)}
\eeq
and by 
\beq
{\cal Y}_2^R= \frac{P_1}{P_2} \frac{m_tm_sV_{31}V_{12}V_{33}^{\dagger}}
{m_cm_bV_{21}V{22}V_{31}^{\dagger}}
\frac{{\cal Q}(\tilde \tau;\tilde t;\tilde W)}
{{\cal F}(\tilde c;\tilde b;\tilde W)+{\cal F}(\tilde c;\tilde \tau;\tilde W)}
\eeq
where $\cal Q$'s are defined as follows:
\beq
{\cal Q}(\tilde \tau;\tilde t;\tilde W)=-32\pi^2i(\frac{m_{\tau}}
{\sqrt 2M_W\cos\beta})\int\sum_{A=1,2}\Delta_{\tau}^RU_{A2}^*
[\epsilon_t\Delta_{\tilde t}^RS_{A2}^*
-\Delta_{\tilde t}^{RL}S_{A1}^*]\tilde G_A
\eeq 
The dressing loop integrals can be expressed in terms of the 
basic integral
\beq
f(\mu_i,\mu_j,\mu_k)=-16\pi^2i\int \Delta_i\Delta_j\tilde G_k
\eeq
 where 
  \beq
f(\mu_i,\mu_j,\mu_k)=\frac{\mu_k}{\mu_j^2-\mu_k^2}
[\frac{\mu_j^2}{\mu_i^2-\mu_j^2}ln\frac{\mu_i^2}{\mu_j^2}-
\frac{\mu_k^2}{\mu_i^2-\mu_k^2}ln\frac{\mu_i^2}{\mu_k^2}]
\eeq
In the limit of no CP violation the analysis limits 
 correctly to the previous results which do not  include
CP violating effects. 

\section{Numerical effects of CP violating phases on proton decay}
We discuss now the numerical size of the effects of CP phases on
the proton lifetime. As is obvious from our discussion above
the proton lifetime is highly model dependent. Specifically
there are two main factors  that govern the lifetime of 
the proton. One of these depends on 
the nature of the GUT sector, i.e.,  if the GUT group is SU(5), 
SO(10), E$_6$,..etc and on the nature of the GUT interaction, e.g.,
on the GUT Higgs structure, while  the second factor that controls
proton decay is the sparticle spectrum and the sparticle interactions
that enter in the dressing loop integrals. 

 In the simplest SU(5) GUT model with two Higgs multiplets $H_1$
 and $H_2$, GUT physics enters mainly via the Higgs triplet mass 
$M_{H_3}$. The proton decay lifetime
is significantly affected if we were to change the GUT structure.
Thus, for example, if one had many Higgs triplets\cite{multi},
 $H_i$ and
$\bar H_i$ (i=1,..,n) where only the triplets $H_1$ and $\bar H_1$
couple with matter, i.e., 
\beq
W_3^{triplet}= \bar H_1J+\bar JH_1+ \bar H_iM_{ij}H_j
\eeq
then the effective interaction on eliminating the Higgs triplets
is  
\beq
W_4^{eff}=-\bar J(M^{-1}_{11})J
\eeq
 Here one finds that the effective Higgs triplet mass is 
$M_{H_3}^{eff}= ((M^{-1})_{11})^{-1}$. 
The model with $M_{H_3}$ and $M_{H_3}^{eff}$ would have very similar  
CP violating effects for the same low energy sparticle spectrum.
The reason is that the CP effects
are largely governed by the nature of the  low energy physics, e.g.,
the sparticle mass spectrum and the couplings of the sparticles  with
matter. Thus we expect similar size CP violating effects in models 
with different GUT structures but with similar size sparticle spectrum 

To discuss the CP violating effects on the proton lifetime 
it is useful to consider the ratio $R_{\tau}(p\rightarrow \bar\nu +K^+)$ 
defined by  
  \begin{equation}
    R_{\tau}(p\rightarrow \bar\nu +K^+)= \tau (p\rightarrow \bar\nu +K^+)/
    \tau_0 (p\rightarrow \bar\nu +K^+)
  \end{equation}
  Here $\tau (p\rightarrow \bar\nu +K^+)$ is the proton lifetime
  with CP violating phases and  $\tau_0(p\rightarrow \bar\nu +K^+)$ 
  is the lifetime without CP phases. This 
  ratio is largly model independent. Thus most of the model 
  dependent features such as the 
  nature of the GUT or the string model  would  be contained 
  mostly in the front factors  such as the Higgs triplet mass, 
  the quark masses, the $A_S$ and $A_L$ suppression factors all of which
  cancel out in the ratio. Similarly the quantity $\beta_p$ which
  is poorly known cancels out in the ratio as do the KM matrix
  elements.  

    We  analyze  $R_{\tau}$
  under the constraints that CP violating phases obey the 
  experimental limits on the electron and the neutron EDMs. For 
  the electron and for the neutron 
  the current experimental limits are\cite{commins,harris} 
  \beq
  |d_e|<4.3 \times 10^{-27} ecm,~~~ 
  |d_n|<6.3 \times 10^{-26} ecm
  \eeq
We are interested in the effects of large phases on the proton lifetime
and for these  to satisfy the EMD constraints we use the cancellation
mechanism. In Fig.2 we present five cases where for different inputs
the electron EDM is plotted as a function of $\theta_{\mu}$.
An analysis of the neutron EDM for the same input is given in Fig.3.
One finds the cancellation mechanism produces several regions where
the EDM constraints are satisfied. 
 In Fig.4 we give a plot of $R_{\tau}$ for the  same  set of inputs
 as in Figs. 2 and 3. The analysis shows that $R_{\tau}$ is a
 sensitive function of the CP phase $\theta_{\mu}$ and variations  of
 a factor of around 2 can occur. 
 We also note that both a suppression as well as an enhancement
 of the proton lifetime can occur as a consequence of the
 CP violating effects. Interestingly the largest  CP effects
 on $R_{\tau}$ occur here at the points of maximum cancellation
 in the EDMs as may be seen by a comparison of Figs.2, 3 and 4.
   The variations in  $R_{\tau}$ due to the phases 
  arise because of constructive and destructive interference between
  the exchange contributions of 
   chargino 1 ($\chi^+_1$) and chargino 2 ($\chi^+_2$) (see Fig.1). 
   We give an illustration of this
  phenomenon in Table 1.
  The analysis of Table 1 exhibits the cancellation in the imaginary
  part of the amplitude for the decay process 
  $p\rightarrow \bar\nu+K^+$ from  
  chargino 1 and chargino 2, and this cancellation leads 
   to an enhancement in the p lifetime ratio for this case.

 It is possible to promote each of the cancellation points in Figs.2
 and 3 into a trajectory in the $m_0-m_{\frac{1}{2}}$  plane  by 
 scaling upwards  by a common scale transformation\cite{in4} 
 \beq
 m_0\rightarrow \lambda m_0,~~~~
 m_{\frac{1}{2}}\rightarrow \lambda m_{\frac{1}{2}}
 \eeq
 The size  of the sparticle 
 spectrum depends on the scale $\lambda$. In general, the larger the
  value of $\lambda$ the heavier 
 is the sparticle spectrum and 
   correspondingly smaller is the CP effect on the dressing loop as 
demonstrated in Fig.5.
    For the minimal SU(5)
        case one needs a relatively heavy spectrum with some of the
        sparticle masses $\sim 1$ TeV to stabilize the proton which has
        the current experimental limit for the  $p\rightarrow \bar \nu K$
        decay mode of
         $\tau(p\rightarrow \bar \nu K)>5.5\times 10^{32}$ yr\cite{takita}.
        Because of the heaviness of the sparticle spectrum, the
        CP effects for the minimal SU(5) model are typically
        small, i.e., of order only a few percent. Larger 
        CP effects can occur in non-minimal models where one
        has several Higgs triplets. Thus we consider an example
        where one has 
        two pairs of heavy Higgs triplets with the Higgs triplet mass 
        matrix given
        by 
        
        \beq
 \left(\matrix{0 & a \Lambda \cr
        a \Lambda &  M_2}
            \right)
\eeq
Such a structure can arise, for example, in an SO(10) model\cite{so10}
with two ${\bf 10's}$ of Higgs and a
{\bf 45} of Higgs with a superpotential of the type 
$W_H= M_2 {\bf 10_{2H}^2}+\Lambda{\bf 10_{1H}45_H10_{2H}}$. 
After the ${\bf 45}$ of
Higgs develops  a VEV  $<{\bf 45_H}>$=$(a,a,a,0,0)\times i\sigma_2$ one
finds that only one pair of Higgs doublets remain massless  while
the Higgs  triplets ($\bar H_{t1},\bar H_{t2}$) and 
($ H_{t1}, H_{t2}$) have the mass matrix given by Eq.(25). 
In this case one has $M_{eff}=a^2\Lambda^2/M_2$ and  one can
arrange for proton stability even with a light spectrum by an 
adjustment of the parameters $a\Lambda$ and $M_2$.

  Finally we discuss the current uncertainties in the proton 
  lifetime predictions. Uncertainties 
  arise from the errors in the quark masses, in $\beta_p$       
and in the KM matrix elements. The largest source 
of uncertainties
arises from the strange quark mass ($m_s$). 
There are several determinations of
$m_s$: $m_s=193\pm 59$ MeV\cite{prades}, 
$m_s=200\pm 70$ MeV\cite{chet},
$m_s=170\pm 50$ MeV\cite{domin},$m_s=155\pm 15$ MeV\cite{lubi}, 
all evaluated at 
the scale  1GeV. We take for our average  
$m_s=180\pm 50$ MeV. For the charm quark mass ($m_c$) we use 
$m_c=1.4\pm0.2$ GeV\cite{caso} while for the bottom quark mass
($m_b$) we use $m_b=4.74\pm 0.14$ GeV\cite{caso}.
The contributions from the  first generation quarks are small and
are not the sources of any significant uncertainty in the p lifetime. 
The errors in the KM matrix elements are of 
a subleading order for the $\bar \nu K$ mode but are still significant 
enough to be
included. We use the results of Ref.\cite{caso} for the allowed   
ranges of the KM matrix elements.
 For $\beta_p$ we use the result of the lattice gauge 
analysis of Ref.\cite{gavela}.
In Fig.6 we exhibit the error corridor for the proton lifetime 
for the case
(1) of Fig.2 with $M_2/a\Lambda=0.01,M_2=M_G$. One finds that given the 
current errors in the input data the predictions for the proton lifetime 
 has an uncertainty of about a factor of 2 ($1^{1.5}_{-0.5}$) 
 on either side of the mean.
A similar  analysis holds for the cases(2-5) of Fig.2.
        We note that the uncertainties in the predictions of 
        the proton lifetime is of the same order as the size of 
        the CP violating effects. It is for this reason that we 
        choose to exhibit the results of our analysis in Figs.
        4 and 5 in terms of the ratio $R_{\tau}$ since the effects
        of the uncertainties cancel in the ratio. The analysis
        also shows that an improvement in the determination of
        the quark masses and of $\beta_p$ is essential for a 
        more precise prediction of the proton lifetime in 
        supersymmetric unification of the type discussed here.
        The reduction of the error in the prediction of p
        lifetime will also help to define the CP effects on 
        proton decay when such a decay is experimentally observed.
 
        In summary the CP violating
         effects on the proton lifetime are relatively large 
         if the sparticle spectrum entering the dressing loop integrals  
         is relatively light and the CP violating effects 
         get progressively smaller
         as the scale  of the sparticle spectrum entering   
         the dressing loops 
         gets progressively larger. The current experimental limits
         on the sparticle masses allow for a relatively light sparticle
         spectrum, i.e., significantly smaller than 1 TeV. 
         This means that there  exists 
         the possibility of significant
         CP violating effects on the proton lifetime.
         However, the minimal SU(5) model does not support the
         scenario with a light spectrum and thus the CP violating
         effects for the case of the minimal model are small.  
         However, for the non-minimal case proton stability can
         occur even for a relatively light spectrum due to suppression
         from a more complicated Higgs triplet sector. In these
         types of models CP violating effects can be significant.

\section{Conclusion}
In this paper we have investigated the effects of CP violating phases
arising from the soft SUSY breaking sector of the theory on
the proton decay amplitudes. It is found that the CP effects can
increase or decrease the proton decay rates and that the size of
their effect depends sensitively on the region of the parameter space
one is in. Effects  as large as a factor of 2 
are seen to arise from CP violating phases in the part of the 
parameter space investigated and even larger effects in the
 full  parameter space may occur. It is found that the CP violating
 effects in the minimal SU(5) model are typically small since a relatively
 heavy sparticle spectrum is needed to stabilize the
 proton in this case and a heavy spectrum suppresses the CP effects
 in the dressing loop integral. However, significantly larger CP
 effects on the p lifetime are possible in non-minimal models
 with more than one pair of Higgs triplets since in these models
 the proton can be stabilized with a relatively light sparticle 
 spectrum. We also investigated the uncertainties in the p lifetime
 predictions due to uncertainties in the quark masses, in $\beta_p$ 
 and in the KM matrix elements. We find that these uncertainties  
 modify the proton lifetime by a factor of 2 around the mean value.
 The observations arrived at in this analysis would be applicable 
 to a wide class of models, including GUT models and string models 
 with dimension five baryon and lepton number violating operators\\

\noindent
Note Added: After the paper was submitted for publication an improved
limit on $p\rightarrow \bar \nu_{\mu}K^{+}$ mode of 
$\tau(p\rightarrow\bar\nu_{\mu}K^{+})>1.9\times 10^{33}$ yr 
has been reported\cite{totsuka}. The new limit does not affect the
conclusions arrived at in this paper.\\
\noindent
{\bf Acknowledgements}\\ 
 This research was supported in part by NSF grant PHY-9901057\\

\begin{center} 
\begin{tabular}{|c|c|c|}
\multicolumn{3}{c}{Table~1:CP effects on Chargino dressings.  } \\
\hline
 case  & Chargino 1 &  Chargino 2 \\
\hline
  (Re${A_{\nu_{\mu}}^{\tilde W_i}}/{|A_{\nu_{\mu 0}}|}$,
  Im${A_{\nu_{\mu}}^{\tilde W_i}}/{|A_{\nu_{\mu 0}}|}$)
  &$(-.22,-.89)$ & $(-.026,.2)$  \\
\hline
\hline
 case  & (Re${A_{\nu_{\mu}}}/{|A_{\nu_{\mu 0}}|}$,
  Im${A_{\nu_{\mu}}}/{|A_{\nu_{\mu 0}}|}$)& 
 ${|A_{\nu_{\mu}}|}/{|A_{\nu_{\mu 0}}|}$   \\
\hline
sum 1\&2  &$(-.25,-.69)$&$.74$\\ 
\hline
\end{tabular}\\
\noindent
Table caption: The table gives an analysis of the dressing loop
integrals for dresssings with Charginos 1\&2 (see Fig.1) and their sum 
for the case when $m_0=71$ GeV, $m_{\frac{1}{2}}=148$ GeV, $\tan\beta=2$,
$\theta_{\mu}=1.4$, $\xi_1=0.3$, $\xi_2=1.8$, $\xi_3=0$, where all phases
are in radians. The
analysis shows cancellations in the dressings between Chargino 1
and Chargino 2 for the case with phases. 
\end{center}

\noindent
{\bf Figure Captions}\\
Fig.1: Examples of the dressing of LLLL baryon-number-violating 
dimension five operators by chargino, gluino and neutralino 
exchanges that contribute to the  
proton decay. Cancellation among diagrams such as between
$\chi^+_1$ and $\chi^+_2$ exchanges can lead to an enhancement 
of the proton lifetime. The dressings of the RRRR dimension five
operators is also exhibited.
\\
Fig.2:
Plot of Log$_{10}|d_e|$ vs $\theta_{\mu}$ exhibiting 
cancellations where the five curves correspond to the five
sets of input for the parameters $\tan\beta$, $m_0$, $m_{1/2}$,
$\xi_1$, $\xi_2$, $\xi_3$,  $\alpha_{A_0}$, 
and $A_0$ given by 
(1)2,71,148,
$-1.15$,$-1.4$,1.27,
$-.4$,4 (dotted),
(2)2,71,148,
   $-.87$,$-1.0$,1.78,
   $-.4$,4 (solid),
(3)4,550,88,
.5,$-1.55$,1.5,
.6,.8 (dashed),
(4)4,750,88,
1.5,1.6,1.7,
.6,.8 (long dashed),
and
(5)2,71,148,
.55,1.,1.35,
$-.4$,4 (dot-dashed). All masses are in GeV and all phases are
in radians.\\
Fig.3:
Plot of Log$_{10}|d_n|$ vs $\theta_{\mu}$ exhibiting 
cancellations where the five curves correspond to the five
sets of input for the parameters $\tan\beta$, $m_0$, $m_{1/2}$,
$\xi_1$, $\xi_2$, $\xi_3$,  $\alpha_{A_0}$, 
and $A_0$ as given in Fig.2. \\
Fig.4:
The ratio $R_{\tau}$ of the proton lifetime with phases and 
without phases as a function of $\theta_{\mu}$ for the five cases
given in Fig.2.\\
Fig.5:
The ratio $R_{\tau}$ as a function
of the scaling factor $\lambda$ defined in the text.
The four curves correspond to the four sets
of input for the parameters $\tan{\beta}$,
$\xi_1$, $\xi_2$, $\xi_3$, $\theta_{\mu}$,
$\alpha_{A_0}$ and $A_0$ given by
(1)2,$-1.15$,$-1.4$,1.27,$-1.7$,$-.4$,4 with
$m_0=71$ and $m_{1/2}=148$ for the point of intersection
with $R_{\tau}$ axis (dotted).,
(2)2,$-.87$,$-1.0$,1.78,$-2.15$,$-.4$,4 with $m_0=71$ and
$m_{1/2}=148$ for the first point (solid).
(3)4,.5,$-1.55$,1.5,1.56,.6,.8 with $m_0=550$ and
$m_{1/2}=88$ for the first point (dashed).
(4)4,1.5,1.6,1.7,$-1.56$,.6,.8 with $m_0=750$
and $m_{1/2}=88$ for the first point (long dashed).
All masses are in GeV and all phases are
in radians. All trajectories
satisfy edms constraints. \\
Fig.6: 
Exhibition of the uncertainties in the proton lifetime predictions
due to uncertainties in the input data for case(1) of Fig.2 where
we assumed $M_2/a\Lambda=0.01,M_2=M_G$.\\

\noindent

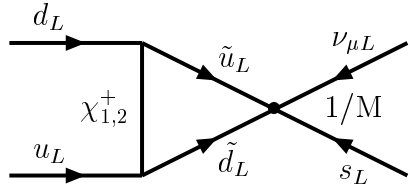
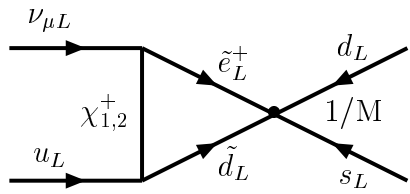
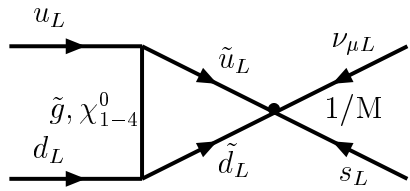
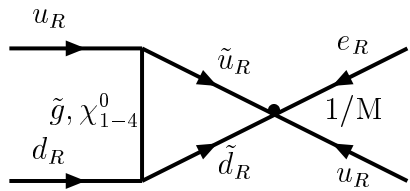
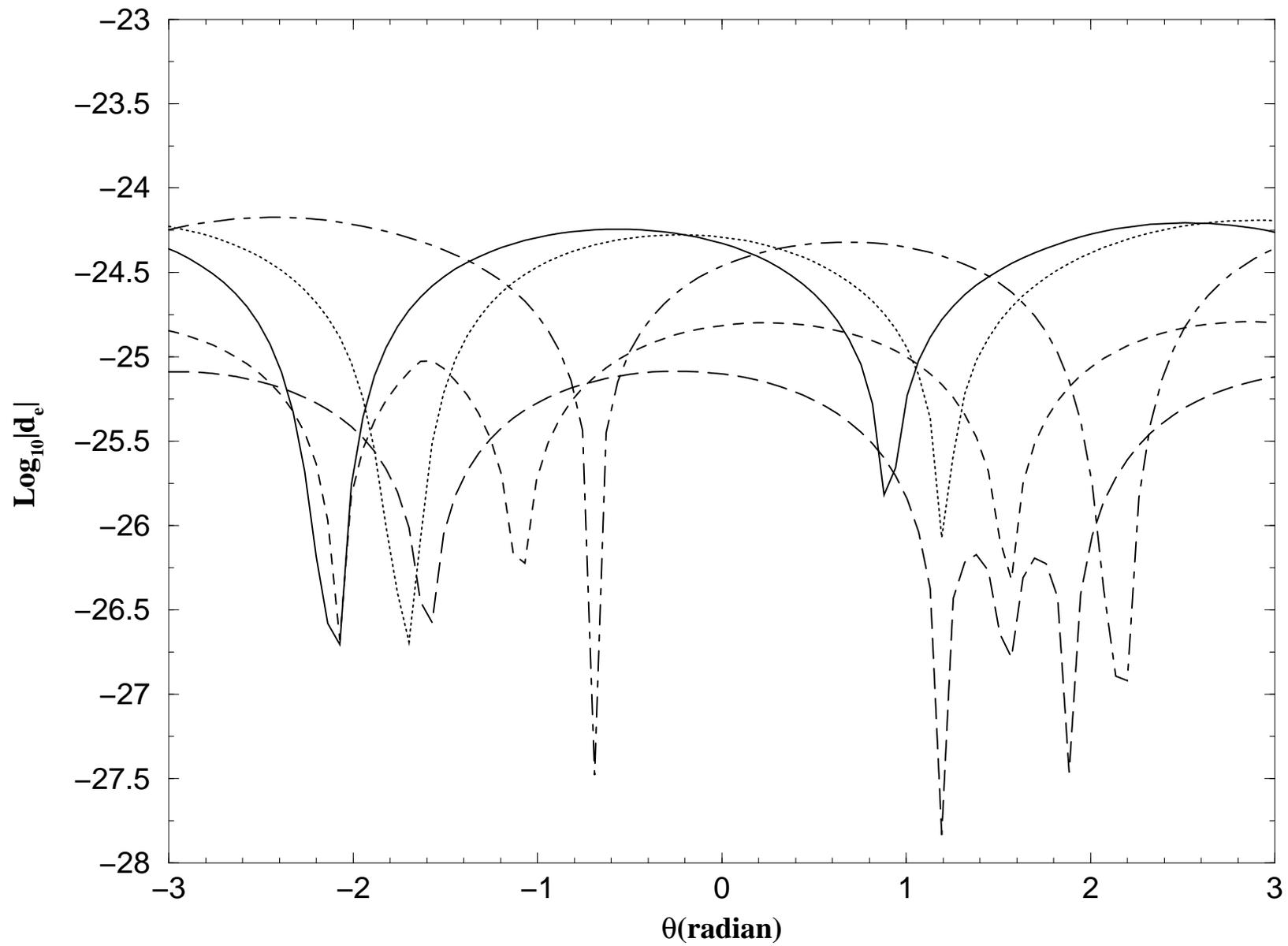
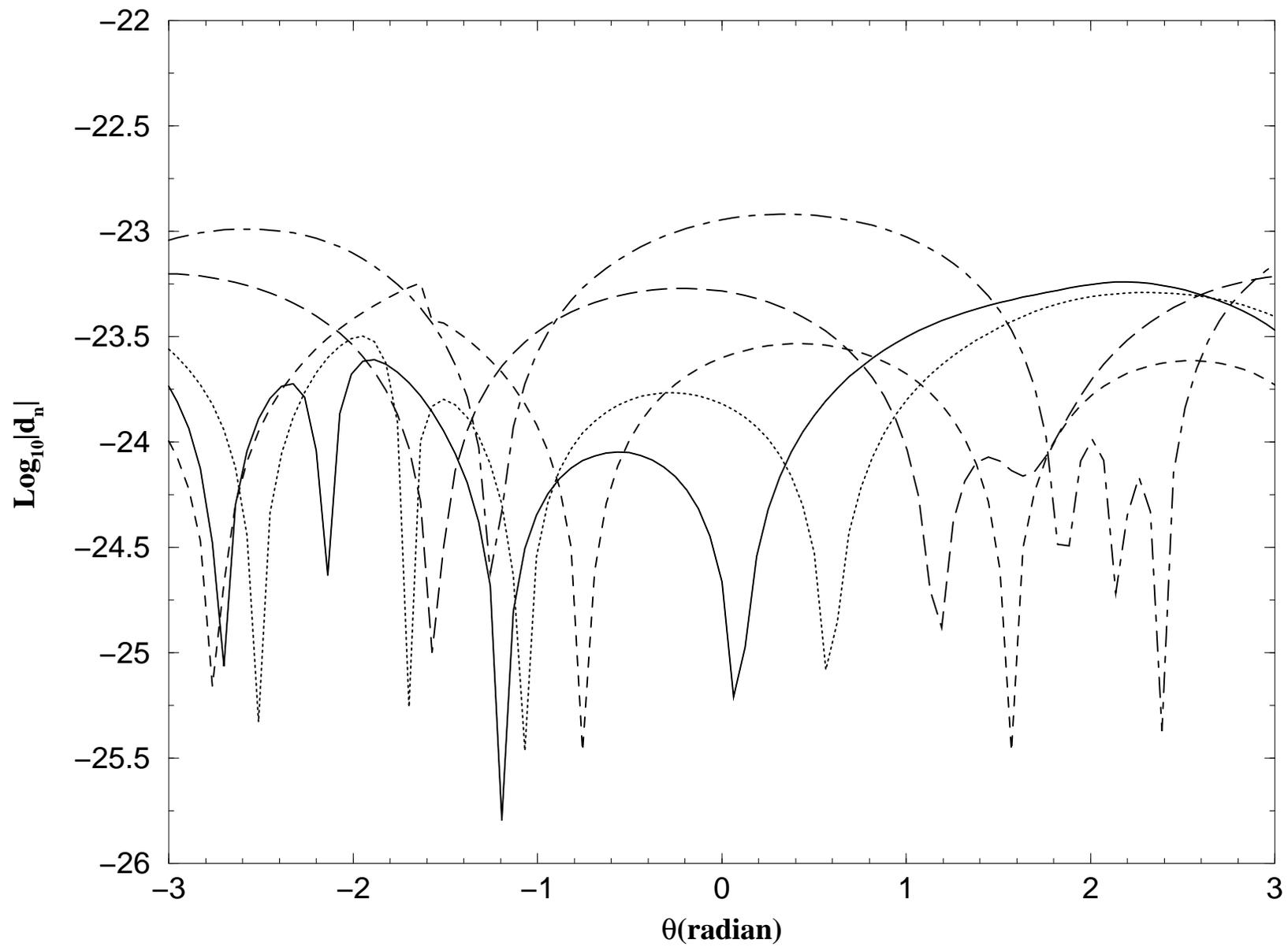
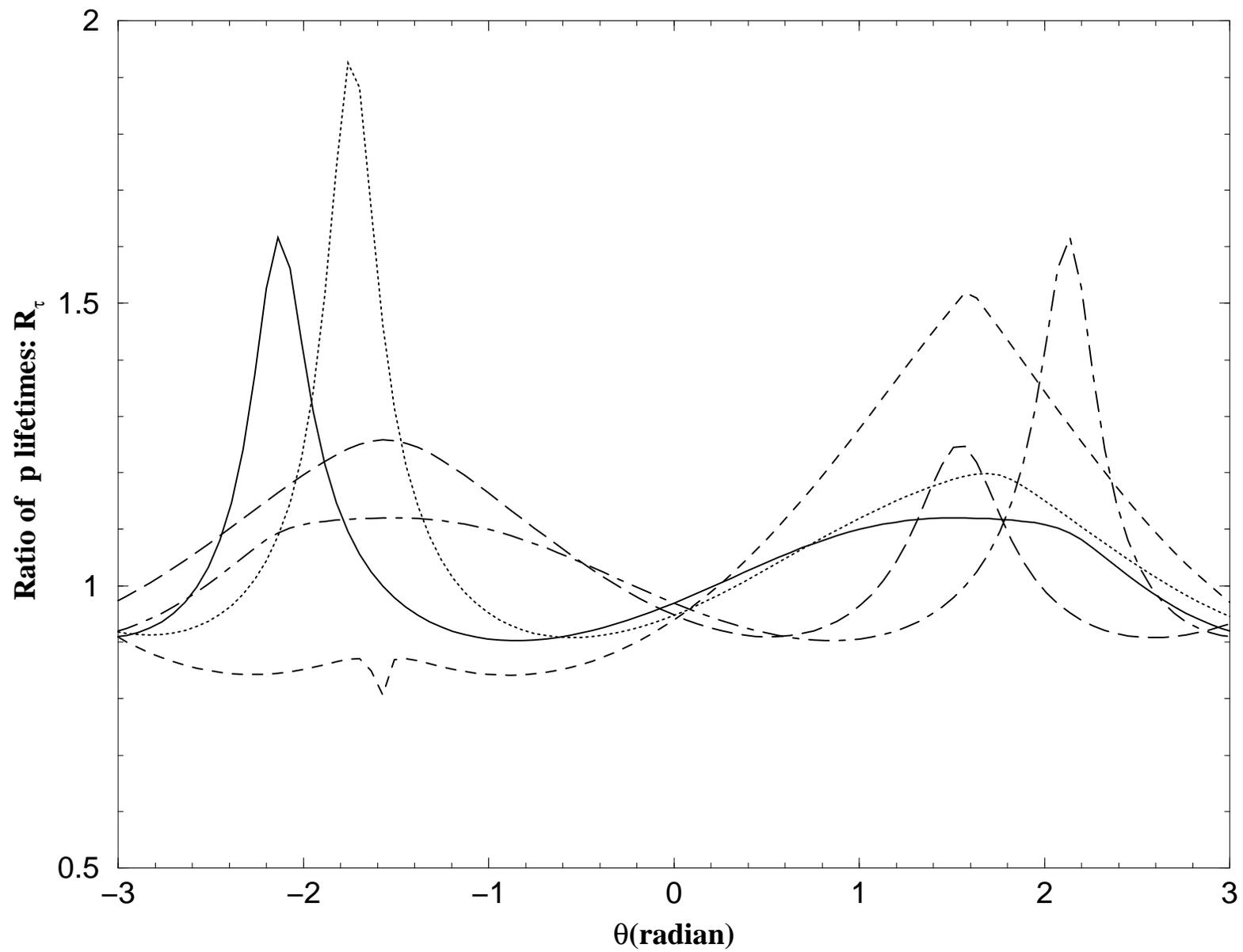
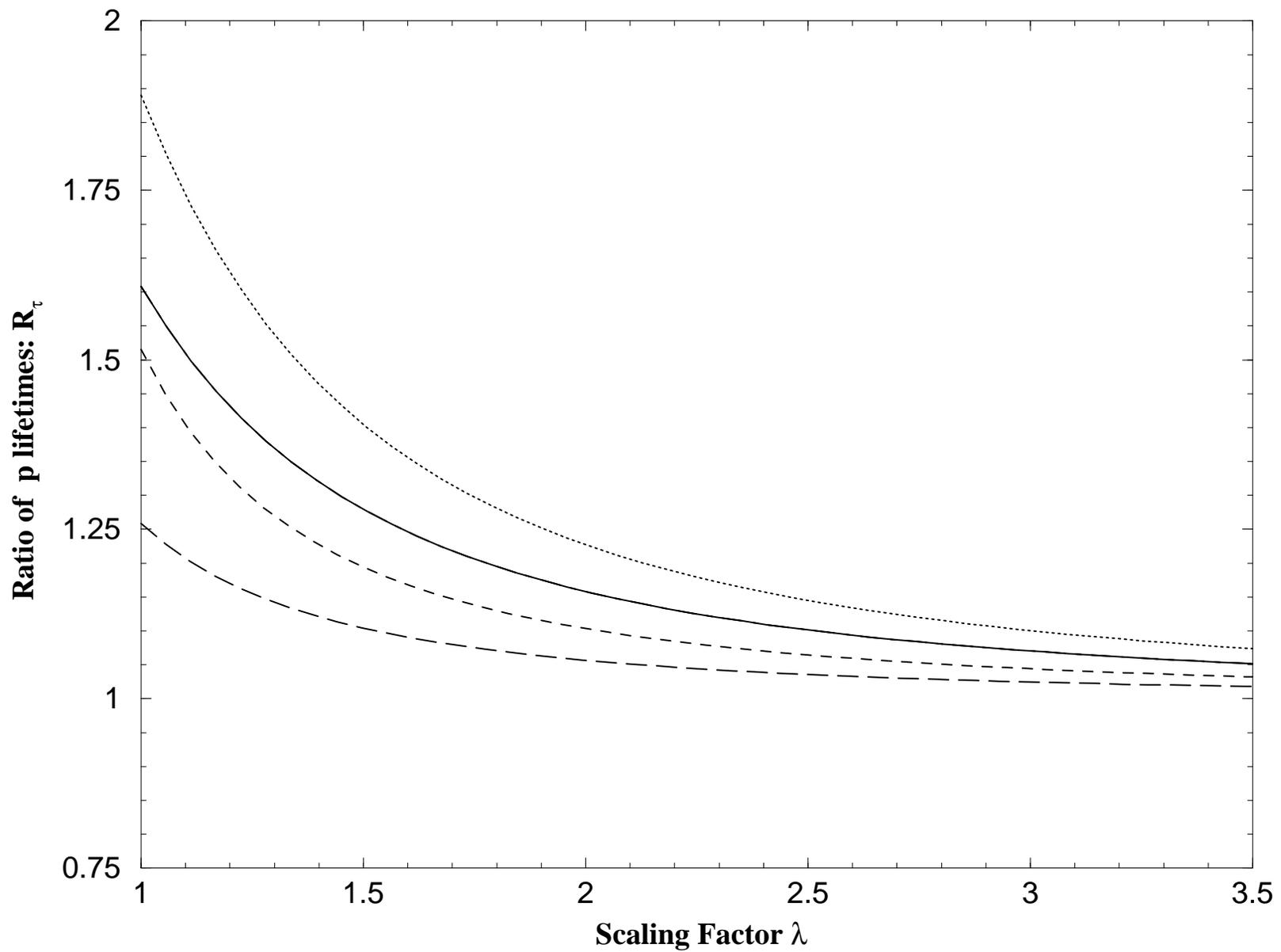
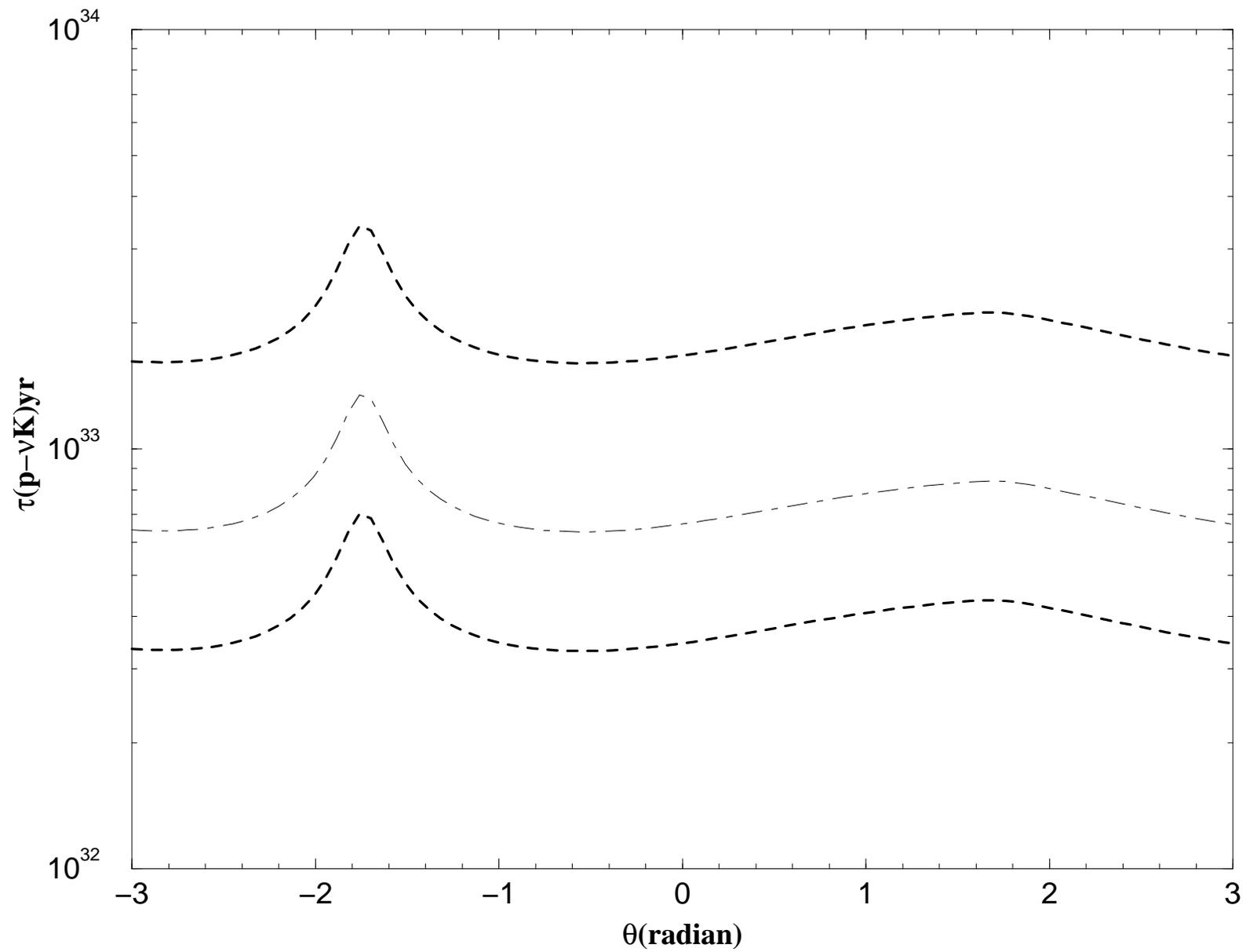
\end{document}